# On Polarization Of The Beam Extracted With The Bent Crystal

Mikhail Ukhanov

*Institute for High Energy Physics*
*Protvino, Moscow region*
*Russia, 142280*

**Abstract.** Particles scattered off the nuclear target acquire a polarization if the nuclei have a non zero analyzing power. This effect is enhanced when particles traverse a bent crystal. Such enhancement under certain assumptions allows one to measure the analyzing power at the level of $10^{-4}$ at the square of transfer momentum $|t| < 10^{-5}$ $(GeV/c)^2$. If it happens that the analyzing power is big enough, then one can get the beam polarization more than 50% after extraction of the primary beam with the bent crystal.

## CONSEQUENCE FROM ANALYZING POWER DEFINITION

### Analyzing Power Definition

In what follows we will always speak about the elastic scattering. Figure 1 depicts the definition of the analyzing power. Analyzing power is the property of the target (or, better to say, the property of the interaction) when a beam of initially unpolarized particles acquires polarization after scattering on the target. Due to the symmetry constraints polarization has opposite sign for the particles scattered to the left and to the right in the scattering plane.

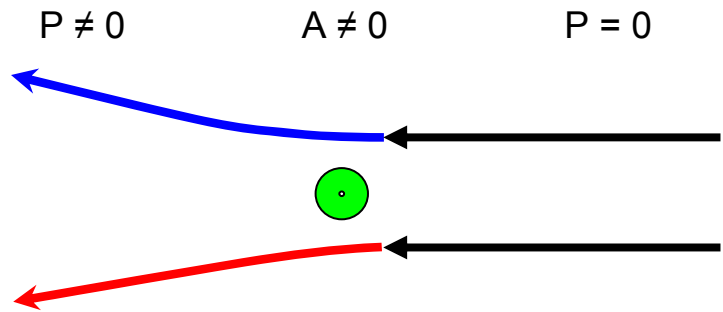

**FIGURE 1.** Definition of the analyzing power. Initially unpolarized particles acquire polarization after scattering. The polarization has opposite sign on the left and right in the scattering plane (paper plane).

## Application to channeling in a bent crystal

If we consider channeling in a bent crystal as a sequence of scatterings all to the same side (e.g. to the left as it is shown in Figure 2), then polarization of the beam will grow after each interaction.

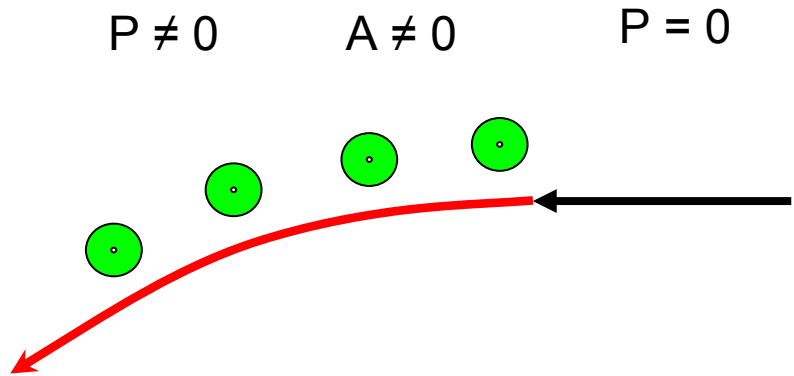

**FIGURE 2.** Channeling in a bent crystal as a sequence of scatterings.

At the exit of the crystal the polarization enhances relatively to the single scattering.

## Consequence

Beam, extracted by means of the bent crystal, must be polarized by definition.

## NUMERICAL EXAMPLE

Polarization of the beam after $N$ scatterings with analyzing power $A$ in each of them can be calculated using the following expression

$$P = \frac{(1+A)^N - (1-A)^N}{(1+A)^N + (1-A)^N}$$

### *Naive Case*

Critical channeling angle in silicon crystal for proton momentum 70 *GeV/c* is equal 20 *μrad* (for any other momentum it is given by 5 *μrad / momentum [TeV/c]*) [1]. Scattering angle of 20 *μrad* results in $N = 5000$ scatterings for 100 *mrad* bending angle. With $A = 10^{-4}$ we get polarization $P = 46\%$.

In this example we considered the channeling proccess as a sequence of identical scatterings all at the same angle (which was taken to be the critical channeling angle). The purpose of this example is to show that even for very small values of the analyzing power amplification in the crystal due to multiple scatterings allows one to get substantial polarization of the extracted beam.

## Less Simplified Example

Now let us allow particles, captured in the channeling, to scatter off the crystal lattice at any angle within the critical angle range. Since the scattering angle varies the analyzing power value cannot be taken as a constant. It is necessary to take it as a function of the transfer momentum *t*. The functional form of *A* versus *t* is model dependent. For very small |*t*| scattering amplitudes can be approximated [2] as $|t|^\alpha f(s)$, where α depends on helicity of particles in initial and final states and *f(s)* is a function of total energy. In this example we adopt linear dependence of the analyzing power on the scattering angle

$$A = A_0 \times \sqrt{\frac{t}{t_0}}$$

where $-t_0 = 10^{-3}$ (GeV/c)$^2$, $|t| < |t_0|$ is the transfer momentum in each scattering and $A_0$ – is the reference value of the analyzing power at $t = t_0$.

Since *A* is proportional to the scattering angle all details of the channeling process cancel out and the final polarization depends only on the crystal bending angle. In other words in this case we do not need to know the exact particle trajectory inside the crystal.

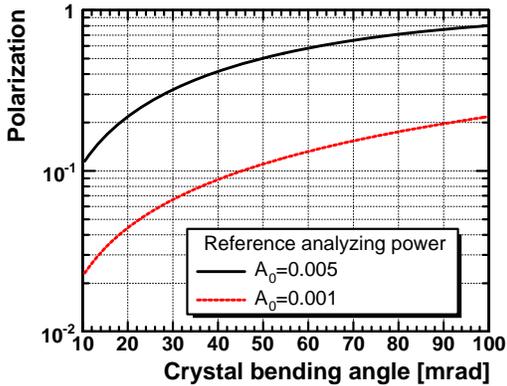 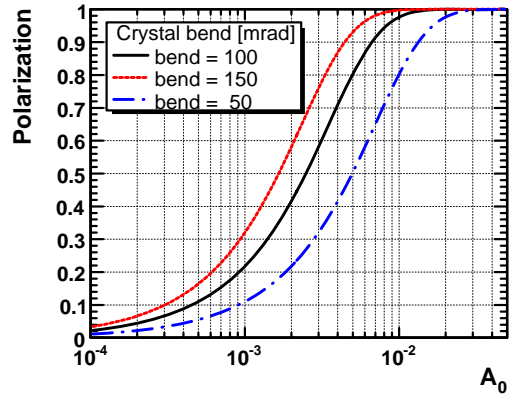

**FIGURE 3.** Dependence of polarization on the bending angle.

**FIGURE 4.** Polarization versus analyzing power reference value for different crystal bending angles.

Figure 3 shows the polarization value for a range of crystal bending angles. If the reference value $A_0$ is big enough (e. g. 5·10$^{-3}$) then the beam deflected by 100 *mrad* can have polarization as high as 80%. It is worth to note that the extraction efficiency by means of the bent crystal changes with the crystal bend. [3] For the bend of 20 *mrad* the efficiency for protons at 70 GeV/c is more then 0.8 while for 86 *mrad* bend it is about 10$^{-4}$.

Figure 4 shows the polarization versus reference value $A_0$ for different crystal bending angles. On this graph we can see that polarization of the extracted beam can serve as a tool to measure small values of $A_0 < 0.001$. Such values of the analyzing power are hardly, if not at all, reachable by any other methods [4].

*General Remarks*

It is evident, that in general case, when the analyzing power dependence on the scattering angle is not linear, calculated value of the polarization of the extracted beam will depend on the assumptions about particles trajectories in the crystal.

Whatever the assumptions are it would be worthwhile to make measurements since no any experimental data exists so far. It is especially intriguing to measure the polarization of heavy ion and antiproton beams deflected by a bent crystal.

## CONCLUSIONS

Polarization effect is enhanced when particles traverse a bent crystal.

Such enhancement allows one to measure the analyzing power at the level of $10^{-4}$ at $|t|<10^{-5}$ $(GeV/c)^2$ provided linear dependence of the analyzing power on the scattering angle.

With this assumption if it also happens that the reference analyzing power $A_0$ is big enough (>0.5%), one can get the beam polarization more than 50% after extraction of the primary beam with the bent crystal what becomes suitable for the practical needs.

## ACKNOWLEDGMENTS

It is my pleasure to thank the SPIN-2006 Organizing Committee for the support and hospitality and IHEP directorate for the support of my trip to Japan. I am also grateful to my colleagues Chujko B.V., Kharlov Yu.V. and Semenov P.A. for the fruitful discussion of the topic of this talk.

This work was supported by RFBR grants 04-02-16381-a and 06-02-28309-z.